\documentstyle[12pt,epsfig]{article}
     
\begin{document}

                 \noindent
\centerline{{\large \bf ON THE ORIGIN OF GAMMA RAY BURSTS}} 

\medskip
\noindent
\centerline{{\bf Nir J. Shaviv and Arnon Dar}}

\medskip
\noindent
\centerline {Department of Physics and Space Research Institute}
\centerline {Israel 
Institute of Technology, Haifa 32000, Israel.}
\medskip
\noindent
{\bf  We propose that repeated
photoexcitation/ionization of high Z atoms of highly relativistic  
flows by star-light in dense stellar regions followed by emission of   
decay/recombination photons, which are beamed and boosted to 
$\gamma$ ray energies in the observer frame, 
produce gamma ray bursts (GRBs). We show that this overlooked  
mechanism is able to convert efficiently baryonic kinetic energy 
release in merger or accretion induced collapse of neutron stars into 
cosmological GRBs and reproduces remarkably well all 
the main observed properties of GRBs.}

The origin of gamma ray bursts (GRBs), 
which have been discovered 35 years ago,  
is still a complete mystery [1]. Their observed isotropy in the sky, 
deficiency of faint bursts and the lack of concentration towards the Galactic 
center, in the Galactic disk and in the direction of M31, strongly 
suggest [2] that they are cosmological in origin [3]. A cosmological origin 
implies [4] that GRBs have enormous luminosities during short periods of 
time, \begin{equation}
  L\approx 10^{50}d_{28}^2\chi\phi_6 \Delta\Omega~erg~s^{-1},
\end{equation}
where $d=d_{28} 10^{28}~cm$ is their luminosity distance, 
$\phi=10^{-6}\phi_6~ergs~cm^{-2}s^{-1}$ 
is their measured energy flux, $\chi$ is a bandwidth
correction factor and $\Delta \Omega$ is the solid angle into which their
emission is beamed. Moreover, the short durations of GRBs imply very compact
sources. Relativistic beaming is required then, both in order to explain the
absence of accompanying optical light and X-ray emission, and in order 
to avoid self opaqueness due to $\gamma\gamma\rightarrow e^+e^-$. All 
these considerations  [4] seem to support the favorite cosmological model 
of GRBs; relativistic  fireballs [5] formed by mergers of neutron stars 
(NS) or neutron stars and  black holes (BH) in close binary systems [6] 
due to gravitational wave emission [7], or by accretion induced collapse 
(AIC) of NS and white dwarfs (WD) [6,8]. Indeed, the above considerations 
and the observed rate of GRBs favor NS mergers/AIC as the source of  
cosmological GRBs. Nevertheless, no mechanism has been convincingly 
shown to be able to convert a large enough fraction of their binding 
energy release into $\gamma$ rays and to explain simultaneously the 
observed complex light curves, duration distribution and spectral 
behavior of GRBs [4]. 

Most of the binding energy released in NS mergers/AIC is expected to be 
in the form of neutrinos, gravitational waves and kinetic energy of 
ejected material [9]. No mechanism is known which converts efficiently 
gravitational waves into $\gamma$ rays. Neutrino annihilation [6] 
($\nu\bar{\nu} \rightarrow e^+e^-$), and neutrino pair production in 
strong magnetic fields ($\nu\gamma_v\rightarrow \nu e^+e^-$) near 
merging/collapsing NS cannot convert enough binding energy into 
$e^+e^-\gamma$  fireballs or relativistic $e^{\pm}$ beams which can 
produce cosmological GRBs, because of baryon contamination [10] 
and the low efficiency of these processes [11]. Thus, if 
NS-NS and NS-BH mergers, or AIC of WD and NS, 
produce GRBs, the production must proceed through conversion of
baryonic kinetic energy of a highly relativistic flow (a fireball  
or a jet) into $\gamma$ rays. 
In this letter we propose that repeated
photoexcitation/ionization of the highly relativistic atoms 
of the flow by star-light in dense stellar regions followed by emission of   
decay/recombination photons which are beamed and boosted to 
$\gamma$  ray energies in the observer frame (see Fig. 1), 
produce cosmological GRBs. We show that this simple mechanism, which 
has been overlooked, is able to convert enough baryonic kinetic energy 
of highly relativistic flows in dense stellar regions into cosmological 
GRBs. We also show that it predicts remarkably well the main observed 
properties of GRBs; their burst size, duration distribution, complex 
light curves and spectral evolution.  

The surface of neutron stars is believed to consist of iron-like nuclei 
[9]. Therefore, we assume that the ejected mass in NS 
mergers or in AIC of NS or WD contains such high
Z nuclei. In view of the uncertainties in modeling merger/AIC  
of compact stellar objects [11], rather than relying on  
numerical simulations, we deduce the total relativistic kinetic 
energy release in such events from observations of type II supernova 
explosions, which are driven by gravitational core collapse into NS or BH. 
In type II supernova explosions, typically, $10M_\odot$ are accelerated 
to a  final velocity of $v\sim  10^4 ~km~s^{-1}$ [9], i.e., to a total final  
momentum $P\sim10 M_\odot v$. Since core collapse is not affected directly by 
the surrounding stellar envelope, we assume that in merger/AIC 
collapse of NS or WD a similar impulse, $\int Fdt\approx P$, is imparted 
to the ejected mass.  If the ejected mass is much smaller than a solar 
mass,  $\Delta M \ll M_\odot$, 
then it is accelerated to a highly relativistic velocity and its kinetic 
energy is given approximately by $E_K \sim Pc\sim 6\times 10^{53}~erg$. 
More than $10^{-3}$ of this highly relativistic baryonic kinetic 
energy must be converted into $\gamma$ rays in order to produce a typical
cosmological GRB of $\sim 10^{51}~erg$. 

The natural birth places of close binary systems are the very dense 
stellar regions in galactic cores and collapsed cores of globular 
clusters (GC). These GC regions have very large photon densities and 
column densities, $n_\gamma\sim 3L/4\pi c R^2 \epsilon~$ and 
$N_\gamma\sim L/\pi c R \epsilon~,$ respectively. For instance, 
the core of our Milky Way (MW) galaxy has a surface brightness, 
$\Sigma\sim 1.2\times 10^7L_\odot~pc^{-2}$, typical photon energies,  
$\epsilon\sim 1~eV$, and a radius, $R\sim 0.5~pc $ [12]. These values 
yield, on average, $N_\gamma\sim 10^{23}~cm^{-2}$. 
(Actually, this value  
underestimates the column density because we have neglected the   
contribution from stars beyond $R$. Its inclusion yields $N_\gamma$  
that are several times larger. A similar value is obtained for the 
nearby M31 galaxy. Much larger values are obtained for active galactic 
nuclei). The highly relativistic ejecta expands, cools and becomes 
transparent to its own radiation after a relatively short time [5]. 
Its partially ionized, highly relativistic atoms see 
the interstellar photons blue shifted to X-ray energies. These photons  
photoexcite/photoionize the high Z atoms of the flow  
which subsequently decay/recombine radiatively by isotropic X-ray 
emission in the flow's rest frame. For a highly relativistic flow 
with a Lorentz factor $\Gamma \gg 1$, the isotropic X-rays are beamed along 
the flow, and as long as $\theta< 1/\Gamma $ 
their energies, $\epsilon_{X}$, are boosted to $\gamma$ ray 
energies, \begin{equation}
\epsilon_\gamma\approx 2\Gamma \epsilon_{X}/(1+\Gamma^2\theta^2),
\end{equation} 
where $\theta$ is the angle of the 
flow direction relative to the observer. The differential 
fraction of the emission that is directed towards the observer  
is given by 
\begin{equation}
dI/d(cos\theta) \approx 2\Gamma^2/(1+2\Gamma^2\theta^2).
\end{equation} 
 
The total baryonic kinetic energy which is converted into 
beamed $\gamma$ ray emission by repeated excitation/decay  
can be estimated as follows: The total photoabsorption cross section of 
an atom in an atomic state $n$ into a state $n'$ is 
\begin{equation}
  \int \sigma_{\nu nn'}d\nu={\pi e^2\over m_ec}f_{nn'},
\end{equation}
where the integral is over the natural line width and where the oscillator
strength $f_{nn'}$ satisfies the sum rule, $\sum_{n'} f_{nn'}=Z$, with Z being
the number of electrons in the system [13]. If the typical energy of the
interstellar photons is $\epsilon \sim 1~eV$, and if their energy is 
boosted in the flow's rest frame by a Lorentz factor $\Gamma\sim 10^3$, 
then their typical photoabsorption cross section by Fe atoms is
\begin{equation}
 \bar{\sigma}\equiv {\int \sigma_{\nu nn'}hd\nu\over \Gamma \epsilon}
={\alpha\over 2} {h\over m_ec}{hcZ\over\Gamma \epsilon}
 \approx 3\times 10^{-18}cm^2.
\end{equation}
This cross section is larger by about seven orders of magnitude 
than the typical cross sections for production of $\gamma$ rays 
in the interstellar medium by 
inverse Compton scattering, bremsstrhalung and synchrotron emission by 
electrons, or by $\pi^0$ production by hadrons  
followed  by $\pi^0\rightarrow 2\gamma$ decay. 
Thus, if the ionized or excited atoms recombine/decay radiatively 
fast enough, then
the typical energy of GRBs (burst size) from NS mergers/AIC  
in GC is given approximately by
\begin{equation}
 E_\gamma \approx \bar{\sigma}N_\gamma(\epsilon_{X}/ m_{A})  
 E_K \approx 3\times 10^{51}~erg,        
\end{equation} 
where $m_{A}\approx 56m_p$.
Note that the burst size is proportional to the total kinetic energy 
$E_K$ of the relativistic flow, and to the column density of the 
radiation field in the GC along the line of sight to the explosion, 
provided 
that the flow is not strongly attenuated by the radiation field. (If NS 
merger/AIC occurs near a bright active 
galactic nucleus, where $L\sim 10^{12}L_\odot$ and $R\sim 10^{17}~cm $, 
i.e., $N_\gamma\sim 10^{29}~ cm^{-2}$, then a large enough fraction of 
the kinetic energy of the relativistic flow can be converted into a GRB 
even by inverse Compton scattering [14,15]. However, such a GRB will  
be extremely short and structureless).

Let us show that the relativistic atoms are only partially ionized. The 
recombination rate (radiative electron capture) of hydrogen-like atoms into
the ground and excited states is given approximately by [13]
\begin{equation}
 r\approx 4\times 10^{-13}Z^2T_{eV}^{-1/2}~cm^3s^{-1},
\end{equation}
where $T_{eV}$ is the temperature in $eV$ in the flow rest frame.
The ionization/excitation rate must adjust itself to the   
recombination/decay  rate. Consequently, in the rest frame of the flow 
\begin{equation}
n_{e11}T_{eV}^{-1/2}\approx c\Gamma n_\gamma\bar{\sigma}\approx
9n_{\gamma5}/\epsilon_{eV},
\end{equation}
where $n_e=n_{e11}\times 10^{11}$, $n_{\gamma}=n_{\gamma5}\times 10^5$ and
$\Gamma=\Gamma_3\times 10^3$. Moreover, in its rest frame the relativistic flow
expands against the external radiation field until its internal pressure equals
the external pressure, i.e.,
\begin{equation}
n_ekT\approx\Gamma^2n_{\gamma}\epsilon~~or, ~~n_{e11}T_{eV}\approx
\Gamma_3^2 n_{\gamma5}\epsilon_{eV}.
\end{equation}
Consequently, $T_{eV}\approx [\Gamma_3\epsilon_{eV}/3]^{4/3}$ and
$n_{e11}\approx 9^{2/3}\Gamma_3^{2/3}/\epsilon _{eV}^{1/3}n_{\gamma5}$. These
rather low temperatures and high densities of the relativistic flows
justify our initial assumption that the inner electronic shells of the
relativistic Fe atoms are not ionized and the radiative 
decay/recombination is fast enough. 

We have constructed a numerical Monte Carlo code which simulates the  
production of cosmological GRBs by highly relativistic flows in dense 
stellar regions [16]. The code employs the quantum mechanical 
cross sections for the relevant photo excitation/ionization and radiative
recombination/decay processes. 
The core of the MW galaxy [12] was  used for modeling the stellar 
environment (density of stars, stellar luminosities and stellar 
temperatures) in a typical GC. Initial distributions of Lorentz factors 
which are consistent with theoretical considerations and observations
have been used. For instance,
shock acceleration usually produces an approximate broken power-law
spectrum of Lorentz factors,
i.e., $dn_{A}/d\Gamma\sim (\Gamma/\Gamma_m)^{-p} $ with $1.5<p<2.5$ 
for $\Gamma\geq\Gamma_m$ and $p<0$ for $\Gamma\leq\Gamma_m$. 
Using our GRB simulation 
code [16] we have found that (a) the main properties of the simulated 
GRBs are not sensitive to fine details and (b) the calculated light curves 
and spectral behavior of simulated GRBs reproduce remarkably well those 
observed in GRBs [1, 17-19]. In particular, the simulated GRBs look
indistinguishable from the observed GRBs. This is demonstrated in Fig. 
2 which compares a simulated
GRB and a GRB  from the BATSE 1B catalog [18] brought in [1] as an
example of a typical complex GRB. Note that the temporal power 
spectra of the observed and the simulated GRB light curves have the 
same universal power-law behavior, $P(w)\equiv \vert\int 
L(t)exp(iwt)dt\vert^2\sim w^{-2}$ [15]. This universal power-law 
behavior can be derived analytically [15,16] from our model. Here we 
summarize briefly approximate  analytical derivations [16] of the other 
main observed properties of GRBs [1,17-19].  
For the sake of simplicity we neglect here general relativistic effects 
(e.g., time dilation and energy redshift) and here that 
the explosions are spherical symmetric and occur at the center of the GCs, 
that the stars within a GC have a uniform spatial 
distribution, the same luminosity and the same effective surface  
temperature, and that the energy flux in the relativistic flow, 
$E^2dn_{A}/dE$, is peaked around a Lorentz 
factor $\Gamma$.  Then our model predicts that:

1. GRB light curves are composed of a smooth background plus 
strong and weak pulses. Strong pulses are  produced when the 
relativistic flow passes 
near stars within the ``beaming cone'', i.e., stars at an angle 
$\theta_*< 1/\Gamma$ relative to the line of sight from the explosion 
to the observer (see eq. 2). When a star is
actually a multiple star system (binary, triplet, etc) the pulse 
becomes a multipeak pulse with very short separation in time 
between the peaks (spikes). Weak pulses are produced by boosting
star-light from stars near the beaming cone. The smooth 
background is produced by boosting the background light in the beaming
cone from all the other stars in the GC. Although the photoexcitation 
of partially ionized atoms produces line emission, the line emission is  
Doppler broadened into a 
continuum by the continuous distribution of the Lorentz factors of the 
atoms in the flow. Thus, the model predicts a continuous energy spectrum 
that at any moment has an approximate broken power-law form (see point 
8) which depends on the energy spectrum of the relativistic 
atoms in the flow and on the spectral properties of the star-light which 
they boost. The ionization state of the atoms cuts off 
emission in the observer frame below  $E_{min}\sim \Gamma_m{\rm I}\sim 
5-25~keV $, where ${\rm I\sim few\times 100~eV }$ is the ionization 
potential of the last bound electrons in the partially ionized atoms [16]. 
It explains why GRBs are not accompanied by detectable X-ray or 
optical-light emission.   

2. The duration of a GRB reflects the spread in arrival
times at the observer of gamma rays produced within the beaming cone 
(we neglect the short formation time of the relativistic flow because the 
dynamical time for NS merger/AIC is much shorter, typically 1ms). It 
is given approximately by 
\begin{equation}
T\sim  R/2c\Gamma^2.
\end{equation}
Hence, relativistic flows with $\Gamma\sim 10^3$ in MW-like
GC ($R\sim 0.5~pc$) produce GRBs that last typically  25 seconds. 
Thus, approximately a one light-year path in a GC is  
contracted into a ten seconds $\gamma$ ray picture in the observer 
frame.     
 
3. The properties of the pulses from single stars
depend on their locations, luminosities and 
spectra, and on the distribution of Lorentz factors in the 
flow.  A pulse from a star at a distance $D_*$  and an angle 
$\theta_*$ begins at a time $t_i\sim D_*\theta_*^2/2c$
after the beginning of the GRB ($t\equiv 0)$. 
Its duration (while its intensity is twice the background) is 
given approximately by 
\begin{equation}
T_p\sim D_*\theta_*d\theta_*\approx b/c\Gamma \sim\L_* R/Lc\Gamma, 
\end{equation}
where $b$ is the impact parameter from the star at which the star's 
light dominates the photon column density of the GC. For main sequence 
stars with $10^{-1}\leq L_*/L_\odot\leq 10^3,$, 
$\Gamma=10^3$ and MW-like GC we obtain $5~ms\leq T_p\leq 50~s$.
For solar like stars $T_p\sim 50~ms$. 
Multiple stars yield multipeak pulses. For instance,
the time difference between the two pulses from a binary star is given by
\begin{equation}
T_b\approx D_*\theta_*d\theta_*/c\approx d_p/\Gamma c, 
\end{equation}
where $d_p$ is the distance between the binary stars projected on the
plane perpendicular to the flow. Thus, $T_b<500(d_{\rm A.U.}
/\Gamma_3)~ms,$ where $d_{\rm A.U.}$ is the binary separation in 
astronomical units. Since a large fraction of the stars are in close binaries,  
triplets, etc, a large fraction of the pulses have a multipeak 
structure. 
 
4. In a GRB from a GC with $N_*$ stars uniformly distributed,    
the average number of stars within the beaming cone 
is $n_p\sim N_*/4\Gamma^2$. Thus, for a MW-like GC and $\Gamma\sim 10^3$,
we expect, on average, $n_p\sim 3$ strong pulses.

5. The average rate of strong pulses in a GRB for a MW-like GC 
is $dn_p/dt\sim cN_*/2R\sim 0.1~s^{-1},$ independent of time. The average 
time-spacing between the strong pulses, $\Delta T$, is given by, 
\begin{equation}
\Delta T=T/ n_p\sim 2R/cN_*\sim 10 s.
\end{equation}
However, the time-spacing between successive pulses is predicted 
to fluctuate considerably around this average time-spacing.

6. If the spectrum of $\Gamma$  in the flow
has a power-law form, $dn_{A}/d\Gamma\sim \Gamma^{-p}$, then  
the energy spectrum of a strong pulse is given approximately by [16]
\begin{equation}
dn_\gamma/dE \sim \bar{\sigma}(\Gamma)(dE/d\Gamma)^{-1}
\Gamma^{-(p-1)}/(\Gamma\theta_*+1/2)
\end{equation}
where $\bar{\sigma}
\sim \Gamma^{-(1+\delta)}$. For $\Gamma\gg 1/\theta_*$ 
$E\sim 2\Gamma \epsilon_{X}$, $\delta\leq 0.5$ one has 
$ dn_\gamma/dE\sim E^{-(p+1+\delta)}.$ 
For  $\Gamma\ll 1/ \theta_*$, one has [16]  
$E\sim \epsilon\Gamma^2$, $\delta\sim 0$, and then 
$ dn_\gamma/dE\sim E^{-(p+1+\delta)/2}.$ Hence the   
spectrum of a pulse has a broken power-law form. The break occurs when  
$2\Gamma\theta_*\sim 1$, i.e., around an 
energy $E_b\approx 2\Gamma\epsilon_{X}\approx \epsilon_{X}/\theta_*~.$
The power-index changes by $\sim (p+1+\delta)/2)\sim
1.5$ from well below the break to  well above the break. Since  
$t_i\sim D_*\theta_*^2/2c$ and $D_*\approx R$, on 
average, $E_b\sim 1/\sqrt{t_i}$ for small $t_i$ and changes to 
$E_b\sim 1/t_i$ for large $t_i$. (For     
a GC with a uniform stellar distribution, the probability of 
$D_*$ is proportional to $D_*^2$ and consequently 
most of the stars have $D_*\approx R$.) 

7. The relative arrival time at a star of atoms with 
a Lorentz factor $\Gamma$ 
is  given by  $t'\equiv t-t_i\approx D_*/2c\Gamma^{-2}$. Such atoms  
boost the star-light to an energy $\epsilon _\gamma\sim 
2 \epsilon_{X}\Gamma$.     
Therefore, the peak energy, 
$E_p\equiv max ~E^2(dn_\gamma/dE)$, decreases during a pulse   
approximately as    
\begin{equation} 
E_p\sim 1/(t'+\delta t_*)^\alpha~; 1/2\leq \alpha\leq 1, 
\end{equation}  
where  $\delta t_*\sim 1-100~ms$ is an added time 
broadening due to the finite time of the explosion and 
the finite size of the region around the 
star where the bulk of the photo excitation/decay takes place.
$E_p$ is maximal right after the pulse begins and decreases 
monotonically afterwards while the photon flux, $dn_\gamma/dE$, peaks 
at a later time which depends on the
distribution of Lorentz factors in the flow. Power-law spectra  
yield photon fluxes during pulses that, 
on average, are time-asymmetric (fast rise and a slower decay) 
with longer pulses being more asymmetric.
Pulses are predicted to be narrower and their 
peak luminosities shifted closer to the beginning of the pulse when 
viewed in higher energy bands (larger Lorentz factors) as demonstrated in 
Fig. 3.

8.  The durations of GRBs have a bimodal distribution which is a trivial 
consequence of their multipulse nature and the fact that, on 
average, $T_p\ll\Delta T$ independent of $\Gamma$: Multipulse GRBs 
have durations which are equal approximately to the sum of the  
time-spacing between their pulses. 
Consequently, $T\sim \Sigma\Delta T\gg T_p$. This produces 
a bimodal distribution which peaks around 
$T_p\sim 0.3~ms$ for single pulse GRBs and 
around $T\sim 25~s$ for multipulse GRBs. 
This is demonstrated in fig. 4 for simulated explosions in a MW-like GC. 
 
9. The photon column densities of MW-like GCs are not large enough to 
attenuate the relativistic flows.  When a relativistic flow emerges from 
a GC it continues to boost star-light into gamma rays outside 
the GC and to emit 
synchrotron (radio) radiation when it traverses interstellar magnetic 
fields. The extended low level $\gamma$ ray emission, which includes 
much broadened and weaker stellar pulses, 
may last for hours. It is, 
however, below the current detection sensitivity of GRB detectors.
The radio emission which lasts for thousands of 
years may be detectable [16]. 
  
10. The relativistic flows may also collide occasionally with interstellar 
gas (stellar winds, planetary nebulae, etc) or with a molecular 
cloud with a sizeable column density in/near the GC. Typical clouds, 
have  $R\sim 5 ~pc$ and $M\sim 10^4M_\odot$, yielding proton column 
densities of $N_p\sim 10^{22}cm^{-2}$. The total inelastic high energy cross 
section of iron nuclei on protons is $\sigma\approx 10^{-24}cm^{-2}$. 
Thus, nuclear collisions of the flow   
with a molecular cloud near the explosion 
will produce pions, and consequently, a burst of $\sim 10^{51}erg$
multi GeV $\gamma$ rays through $\pi^0\rightarrow 2\gamma$ decays
(and neutrinos through $\pi^{\pm}\rightarrow \mu\nu_\mu$ and 
$\mu\rightarrow e\nu_e\nu_\mu$ decays) with a power-law
spectrum $dn_\gamma/dE\sim E^{-p}$. For $p\sim 2$, production of $\sim 20~
GeV$ $\gamma$ rays comes from $\pi^0$'s  produced mainly by nuclei 
with $\Gamma\sim 200$. Thus $\pi^0$ produced  
$\gamma$ rays of $20~GeV$ 
are delayed, typically, by $D/2c\Gamma^2\sim 2h$,    
as was observed in the case of the 17 February 1994 GRB [20].  

In conclusion, we have described a simple mechanism by which
NS mergers/accretion induced collapse in dense galactic cores 
produce cosmological GRBs. The remarkable success of the model in 
reproducing all the main observed temporal and spectral properties of GRBs  
[16] strongly suggests that GRBs are $\gamma$ ray tomography pictures of 
dense cores of distant galaxies. The proposed mechanism may play
an important role also in other astrophysical gamma ray sources such 
as AGN, pulsars and other cosmic accelerators. 

{\bf Acknowledgement}: We thank A. Laor, 
J.P. Lasota and A. Ori for useful comments and 
the Technion fund for the promotion of research for support.
\vfil
\eject
     \parindent 0cm
\centerline{References}

[1] See, e.g., G.J. Fishman and C.A.A. Meegan, Ann. Rev. Astr. 
 Ap. {\bf 33}, 415 (1995).  

[2] See, e.g., M.S. Briggs, 
Ap. \& Sp. Sc.  {\bf 231}, 3 (1995).

[3] V.V. Usov, and G.B. Chibisov, Sov. Astr. {\bf 19}, 115 (1975);
S. van den Bergh, Astr. \& Ap. Suppl. {\bf 97}, 385 (1983).

[4] For a review of cosmological models see C.D. Dermer 
and T.J. Weiler,  Ap. \& Sp. Sc.  {\bf 231}, 377 (1995).

[5] G. Cavallo and M.J. Rees, Mon. Not. Roy. Astr. Soc. {\bf 183}, 359
(1978); B. Paczynski, Ap. J. {\bf 308}, L43 (1986); J. Goodman, Ap. J. 
{\bf 308}, L47 (1986).

[6] J. Goodman, A. Dar and S. Nussinov,  Ap. J. {\bf 314}, L7 (1987);
D. Eichler et al., Nature, {\bf 340}, 126 (1989).

[7] R.A. Hulse and J. H. Taylor, Ap. J., {\bf 368}, 504 (1975).

[8] A. Dar et al., Ap. J. {\bf 388}, 164 (1992); S.E. Woosley, Ap. J. 
{\bf 405}, 273 (1992).

[9] See, e.g., S. L. Shapiro and S.A. Teukolsky,  ``Black 
Holes, White Dwarfs and Neutron Stars'' (A. Wiley Intersc. Pub.) 1983.

[10] B. Paczynski, Ap. J. {\bf 363}, 218 (1990).

[11] H. T. Janka and M. Ruffert, Astr. \& Ap. {\bf 307},
L33 (1996).  

[12] D. A. Allen, ``The Nuclei of Normal Galaxies'' (eds. R. Genzel \& A.I    .
Harris, 1994) p. 293.     

[13] See, e.g., Ya. B Zel'dovich and Yu. P. Raizer, 
``Physics of
Shock Waves and High Temperature Hydrodynamic Phenomena'' (Academic Press 
1967).

[14] A. Shemi, Mon. Not. Roy. Astr. Soc. {\bf 269}, 1112 (1994). 

[15] N.J. Shaviv and A. Dar, Mon. Not. Roy. Astr. Soc. {\bf 277}, 287 
(1995). 

[16] For details see N.J. Shaviv, Ph.D Thesis, 1996 (Technion Report 
Ph-96-1)
 
[17] See, e.g., J.P. Norris, et al., Ap. J. {\bf 459}, 333 
(1996) and references therein.
 
[18] See, e.g.,  D.L. Band et al., Ap.J. {\bf 413}, 231 (1993); L.A. Ford
et al., Ap.J. {\bf 439}, 307 (1995); B.J. Teegarden , Astr. \& Sp. Sc. {\bf 
231}, 137 (1995).

[19] G. Fishman et al., Ap. J. Suppl. {\bf 92}, 229 (1994).
 
[20] K. Hurley et al., Nature {\bf 372}, 652 (1994).

\begin{figure}
\centerline{
\epsfig{file=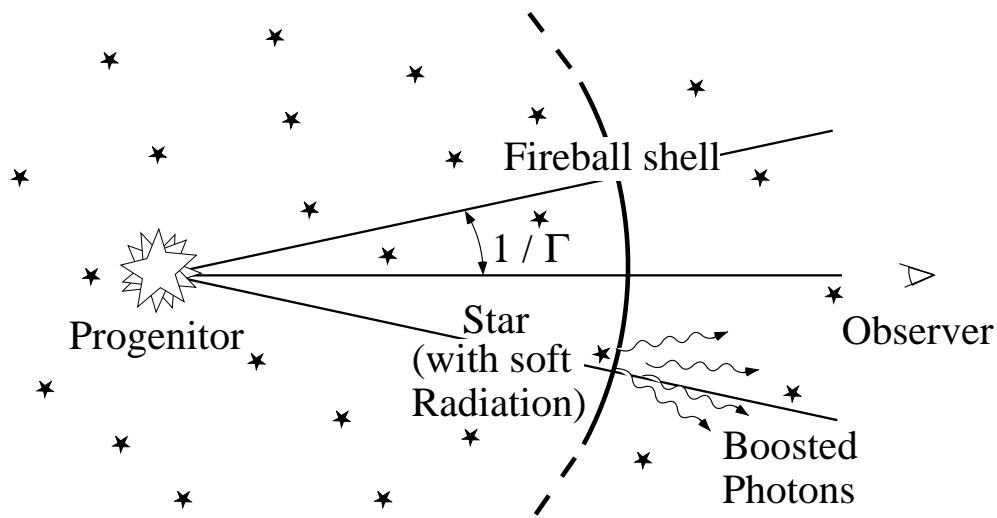,height=3in}}
\caption{A schematic drawing illustrating the formation of a GRB
by a highly relativistic spherical flow in a dense stellar region. Most of
the observed $\gamma$ rays are produced by the radiative decay of
photoexcited atoms near stars within a cone with an opening angle 
$\theta\sim 1/\Gamma$ along the direction to the observer. }
\end{figure}

\begin{figure}
\centerline{
\epsfig{file=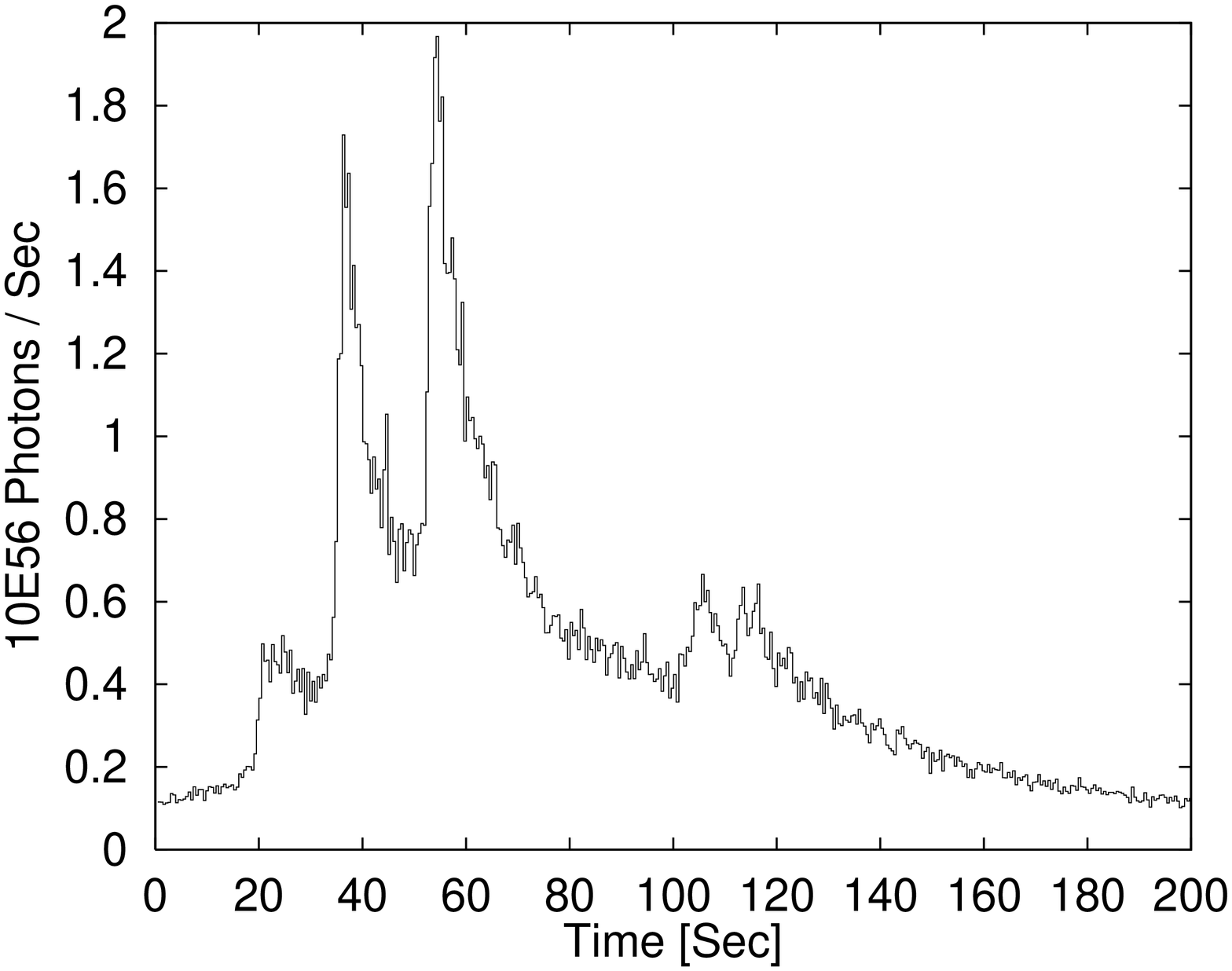,height=2.5in,angle=-90}
\epsfig{file=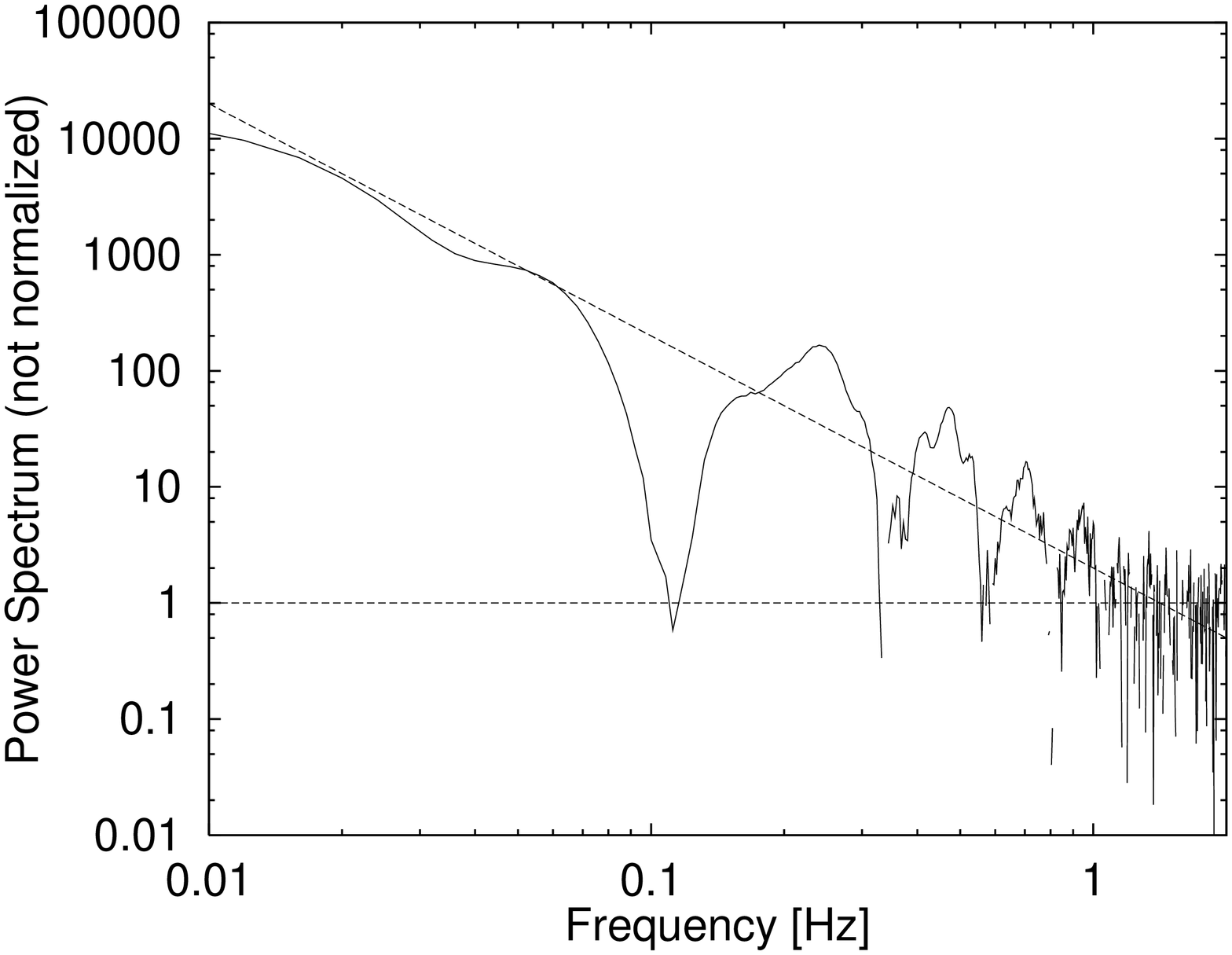,height=2.5in,angle=-90}
}
\centerline{
\epsfig{file=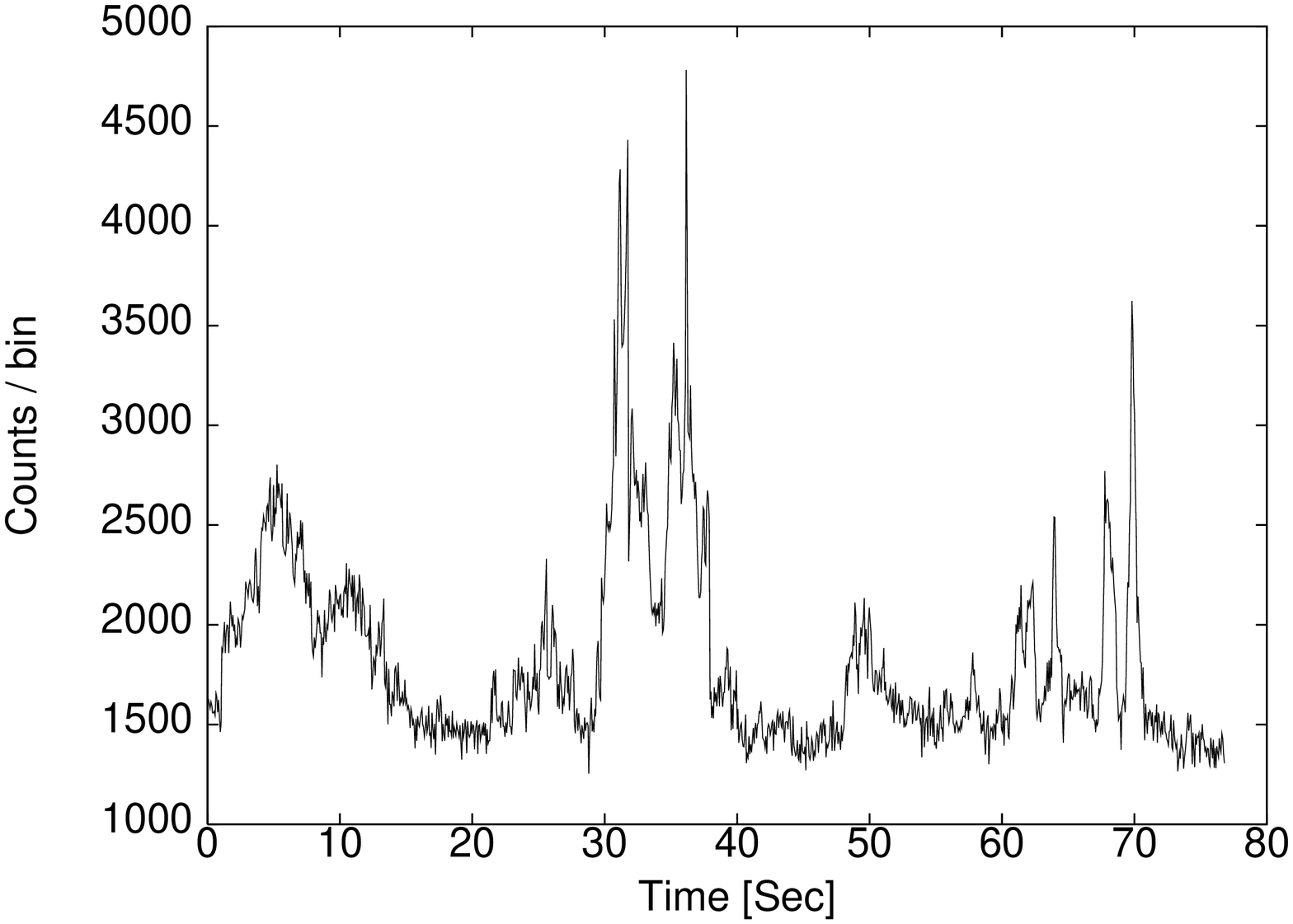,height=2.5in,angle=-90}
\epsfig{file=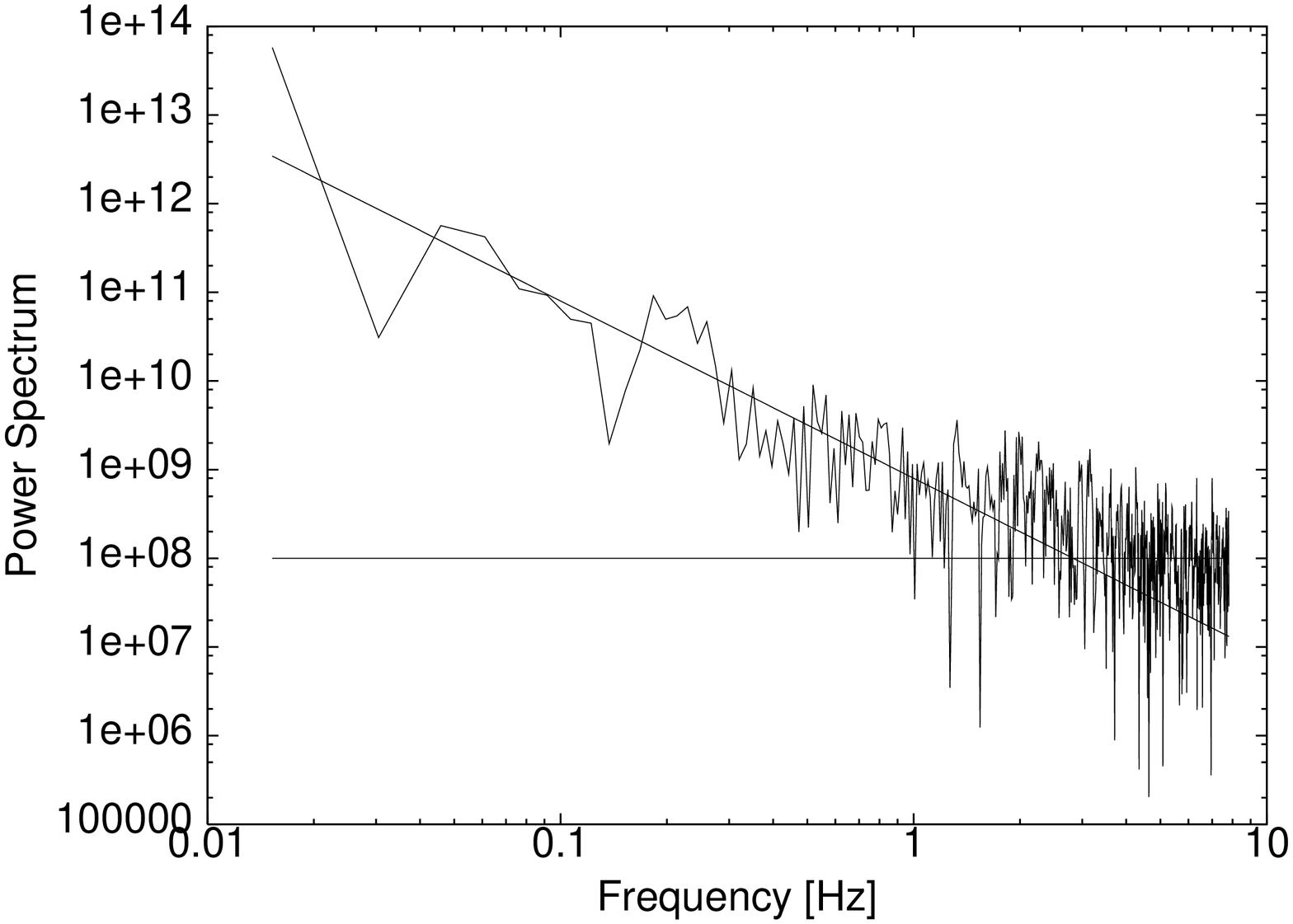,height=2.5in,angle=-90}
}
\caption{(a) A light curve of a simulated GRB and its temporal power 
spectrum. The straight line represents the $w^{-2}$ power law dependence. 
(b)  The light curve of GRB 920110 from the BATSE 1B catalog [18] and its   
temporal power spectrum. }
\end{figure}

\begin{figure}
\centerline{
\epsfig{file=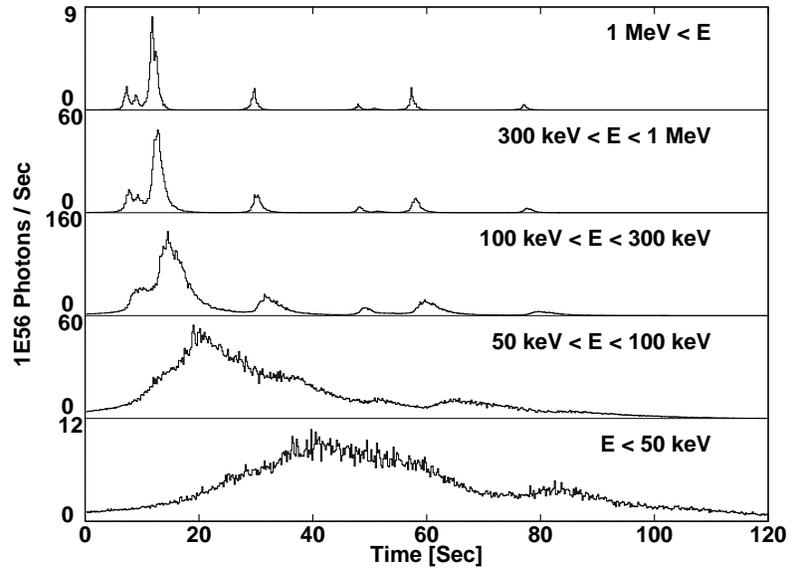,height=3in}}
\caption{ A simulated GRB from a MW-like GC viewed in different energy 
bands.  }
\end{figure}

\begin{figure}
\centerline{
\epsfig{file=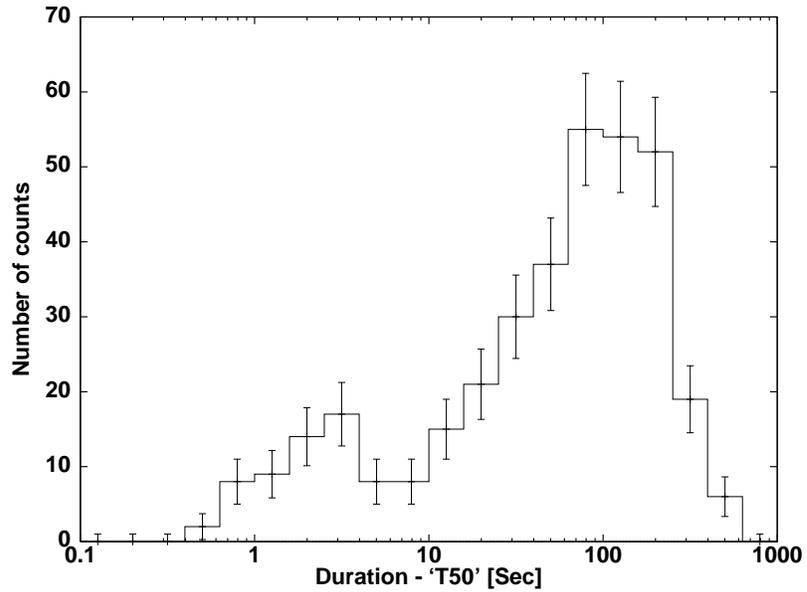,height=3in}}
\caption{The duration distribution of simulated GRBs from a MW-like GC [16].}
\end{figure}

\end{document}